# Terahertz Time-Domain Spectroscopy as a Universal Defect Fingerprinting Tool for Organic Halide Perovskite Solar Cells

**Inhee Maeng**[1,†,*], **Young Mi Lee**[2,†], **Jinwoo Park**[3], **Seng Jae Oh**[1] **and Min-Cherl Jung**[4,*]

[1] YUHS-KRIBB Medical Convergence Research Institute, College of Medicine, Yonsei University, Seoul 03722, Republic of Korea
[2] Beamline Department, Pohang Accelerator Laboratory, POSTECH, Pohang 37673, Republic of Korea
[3] Department of Physics, University of Seoul, Seoul 02504, Republic of Korea
[4] Global Institute of Future Technology, Shanghai Jiao Tong University, Shanghai 200240, China
[†] These authors contributed equally to this work.
[*] Correspondence: inheem@yonsei.ac.kr and mincherl.jung@sjtu.edu.cn

---

**Abstract:** Organic–inorganic hybrid perovskites (OHPs) deliver certified single-junction power conversion efficiencies (PCEs) exceeding 26% and perovskite–silicon tandem values surpassing 34%, yet a substantial gap with the Shockley–Queisser (S–Q) limit persists—primarily due to grain-boundary (GB) defects that drive non-radiative recombination, ion migration, and degradation. Rational passivation demands a non-contact tool capable of identifying and quantifying specific defect species in device-relevant thin films, a capability absent from conventional probes. This short review demonstrates that terahertz time-domain spectroscopy (THz-TDS, 0.3–3.0 THz) fulfills this role. Across four OHP compositions fabricated by sequential vacuum evaporation (SVE)—$MAPbI_3$, $MAPbBr_3$, $FAPbI_3$, and γ-$CsPbI_3$—the THz spectral window captures both intrinsic phonon modes and GB-localized molecular defect vibrations, enabling species-specific, quantitative characterization at room temperature. Notably, the oscillator strength of the SVE-specific 1.58 THz absorption in $MAPbI_3$ scales linearly with XPS-quantified $CH_3NH_2$ defect concentration, establishing THz-TDS as a direct, non-destructive defect meter. Building on these findings, we propose a three-pillar framework for THz-guided defect engineering: (I) quantitative defect measurement via oscillator-strength analysis, (II) material-specific fingerprint identification from a systematically constructed THz library, and (III) fingerprint-guided defect elimination with real-time feedback—together defining a closed-loop quality-control cycle that connects spectroscopic diagnosis to passivation strategy and, ultimately, to enhanced solar-cell efficiency.

---

## 1. Introduction

Organic halide perovskites (OHPs) of the $ABX_3$ crystal family (A = $CH_3NH_3^+$ [MA], $CH(NH_2)_2^+$ [FA], $Cs^+$; B = $Pb^{2+}$; X = $I^-$, $Br^-$) have emerged as one of the most disruptive developments in photovoltaics over the past decade [1]. From a PCE of 3.8% in 2009, certified single-junction OHP solar cells now exceed 26.7% [2], approaching silicon in single-junction configuration and surpassing it in tandem. Longi Green Energy reported a perovskite–silicon two-terminal tandem efficiency of 34.85% in 2025, as certified by NREL [3], exceeding the Shockley–Queisser limit for single-junction devices and underscoring the transformative commercial potential of OHP technology. These advances rest on the exceptional intrinsic properties of OHPs: tunable bandgaps of 1.1–3.0 eV by halide substitution, absorption coefficients exceeding $10^4$ $cm^{-1}$, long carrier diffusion lengths (> 1 μm in thin films, > 175 μm in single crystals), and scalable deposition by solution and vacuum methods [4,5].

Despite these achievements, the gap between laboratory-record efficiencies and the Shockley–Queisser theoretical limit (~30.5% for the ~1.55 eV bandgap of $MAPbI_3$; the absolute S–Q maximum of ~33.7% occurs near 1.34 eV) remains substantial, and long-term operational stability under heat, humidity, and illumination continues to impede commercialization. Both shortcomings converge on a single root cause: defects—particularly those at grain boundaries (GBs) in polycrystalline thin films. GB defects in OHP act as deep-level non-radiative recombination centers, reducing open-circuit voltage ($V^{oc}$) and fill factor (FF) below their radiative limits [6]. Ion migration through GB defect pathways drives phase segregation, halide redistribution, and irreversible degradation under operating conditions [7,8]. Consequently, GB defect passivation has emerged as the central engineering challenge for achieving simultaneously high efficiency and stability in OHP photovoltaics. Recent strategies—including ammonium-salt treatments, sulfonium-based passivators, and ultra-thin polymer interlayers—have extended device lifetimes to over 4,500 hours under continuous illumination [9,10].

Maximizing the impact of defect passivation strategies critically depends on the ability to identify, classify, and quantify the specific defect species present in each OHP composition and fabrication method. Yet, this remains a formidable analytical challenge. Conventional structural probes—X-ray diffraction (XRD) and scanning electron microscopy (SEM)—reveal grain morphology and crystallographic phase but are insensitive to the molecular-level chemical states of GB defects. Photoluminescence (PL) spectroscopy provides defect-density information through non-radiative recombination rates but convolutes multiple recombination pathways and cannot distinguish specific defect chemical species. X-ray photoelectron spectroscopy (XPS) directly identifies defect chemical states with high sensitivity but is surface-limited (~10 nm probing depth), requires high-vacuum sample transfer, and cannot be applied non-destructively to device-integrated films [12]. Impedance spectroscopy accesses trap densities through device-level measurements but requires electrical contacts and cannot resolve individual defect species [13]. Consequently, comprehensive, non-contact, direct characterization of molecular defect species in OHP GB regions—the very defects that limit efficiency—remains an unmet need.

Terahertz time-domain spectroscopy (THz-TDS) operates in the 0.1–3.0 THz range (0.4–12.4 meV), a spectral window that directly overlaps with: (i) low-energy phonon vibration modes of OHP lattices (Pb–X cage vibrations, organic cation librations), (ii) molecular vibrations of defect species at grain boundaries, and (iii) intraband free-carrier dynamics. As a contact-free, non-destructive probe operating at room temperature under ambient conditions, THz-TDS is uniquely suited for direct, species-specific defect characterization. Early optical pump–terahertz probe (OPTP) studies by Ponseca *et al.* [14] confirmed ultrafast, balanced free carriers in $MAPbI_3$ with a sum mobility ($\mu_e + \mu_h \approx 25$ $cm^2$ $V^{-1}s^{-1}$), establishing THz-TDS as a benchmark for OHP carrier quality. Subsequent work revealed phase-dependent phonon mode transformations [11], optical phonon–carrier coupling [15], and electron–phonon interactions [16] using THz spectroscopy. Near-field THz nanoimaging recently demonstrated GB-resolved trap mapping with sub-20 nm spatial resolution [17]. Together, these studies establish THz-TDS as a powerful—yet underexploited—tool for OHP defect diagnostics.

In this short review, we present a comprehensive account of THz defect fingerprints in OHP thin films fabricated by sequential vacuum evaporation (SVE)—a scalable, substrate-universal vacuum deposition method. By systematically mapping THz fingerprints across four OHP compositions—$MAPbI_3$, $MAPbBr_3$, $FAPbI_3$, and $CsPbI_3$—we construct a material-specific THz defect library and show how it enables a three-pillar engineering cycle that closes the loop between defect identification and elimination, positioning THz-TDS as an indispensable quality-control tool for OHP manufacturing.

## 2. Defects in OHP Thin Films: Classification, Origin, and Impact

Defects in polycrystalline OHP thin films originate from multiple sources and manifest in distinct chemical and structural forms. Point defects—including halide vacancies (V_I, V_Br), B-site vacancies (V_Pb), A-site vacancies (V_MA), interstitials (I_i, Pb_i), and antisites—form within grains with formation energies governed by chemical potential [18,19]. Among these, iodine interstitials and MA-on-Pb antisites create deep trap levels (>0.5 eV from band edges) that dominate non-radiative recombination in $MAPbI_3$. Grain boundaries concentrate these point defects due to under-coordinated $Pb^{2+}$ and $X^{-}$ sites at the lattice termination, making GBs the primary recombination-active regions in polycrystalline films [20].

The fabrication method profoundly determines GB defect chemistry. Solution-processed OHP films form GBs during solvent evaporation with residual solvent molecules and incomplete crystallization creating point-defect clusters. Vacuum-deposited films formed by SVE (sequential deposition of MAI and $PbI_2$ in separate steps) create smaller, denser grain structures with higher GB area density. Critically, incomplete perovskite reaction under vacuum conditions at GBs leaves residual $CH_3NH_2$ molecular entities incorporated into the Pb–I bonding network at the GB interface—a process-specific molecular defect class absent in solution-processed films [21].

Beyond chemical point defects, structural discontinuities at phase boundaries introduce lattice strain-mediated defect states. In mixed-phase $FAPbI_3$, the photoinactive δ-phase and photoactive α-phase coexist during thermal processing, creating δ/α interface regions with structurally frustrated Pb–I bonding configurations [22]. Halide compositional non-uniformity at GBs in $FAPb(Br,I)_3$ leads to I-rich and Br-rich domain segregation under operating conditions [23,24]. Characterizing these diverse defect classes across multiple OHP compositions requires a probe with molecular vibrational specificity, which conventional electrical and optical methods cannot provide.

## 3. THz-TDS Methodology and the Lorentz Harmonic Oscillator Analysis

THz-TDS measurements were performed in transmission geometry using a photoconductive antenna-based system. THz pulses were generated and detected by ZnTe(110) electro-optic crystals driven by an 800 nm femtosecond laser (35 fs pulse duration, 80 MHz repetition rate). The spectral range covered 0.2–2.5 THz with dynamic range >40 dB, enclosed in a dry $N_2$-purged chamber to suppress atmospheric water-vapor absorption lines. OHP thin films were deposited on $Al_2O_3$ or sapphire substrates and measured at room temperature without contacts or encapsulation.

The complex optical conductivity $\tilde{\sigma}(\omega)$ was extracted from the measured complex transmission $\tilde{T}(\omega) = \tilde{T}_{sam}(\omega)/\tilde{T}_{sub}(\omega)$ using the thin-film approximation [25]:

$$\tilde{\sigma}(\omega) = [(n_{su\beta} + 1) / (Z_0 \cdot d)] \cdot [1 - \tilde{T}(\omega)] / \tilde{T}(\omega)$$

where $n_{su\beta}$ is the substrate refractive index, $Z_0 = 377\ \Omega$ the free-space impedance, and d the film thickness [26]. To decompose carrier transport and phonon/defect resonances, a multi-oscillator Lorentz harmonic oscillator (LHO) model was fitted [21,27]:

$$\tilde{\sigma}(\omega) = \varepsilon_0 \cdot \Sigma_j \left[ \Omega_j^2 \gamma_j i\omega / (\omega_j^2 - \omega^2 + i\gamma_j\omega) \right]$$

where $\omega_j$, $\Omega_j$, and $\gamma_j$ are respectively the resonance frequency, oscillator strength, and scattering rate of the j-th mode. Defect modes are distinguished from intrinsic phonon modes by: (i) sensitivity to post-annealing treatment, (ii) correlation with XPS-quantified chemical defect concentration, and (iii) absence in solution-processed reference films.

## 4. THz Defect Fingerprint Library Across OHP Compositions

### 4.1. $MAPbI_3$: The 1.58 THz Fingerprint of $CH_3NH_2$ Grain-Boundary Defects

The most definitive example of a process-specific THz defect fingerprint was established in 2019 by Maeng *et al*. [21], who showed that SVE-fabricated $MAPbI_3$ thin films exhibit a strong, narrow absorption at 1.58 THz with an absorption coefficient exceeding 11,000 $cm^{-1}$—approximately two orders of magnitude above the background phonon level. Figure 1 compares the THz transmission spectra of $MAPbI_3$ fabricated by SVE, two-step solution, anti-solvent, and the $MASnI_3$. Only the SVE sample shows the 1.58 THz absorption, accounting for ~50 % suppression of THz transmission at that frequency. LHO fitting resolves three oscillators at 0.95, 1.58, and 1.87 THz, where the 0.95 and 1.87 THz modes are present in all samples as intrinsic lattice phonons, while the 1.58 THz oscillator is exclusive to SVE fabrication.

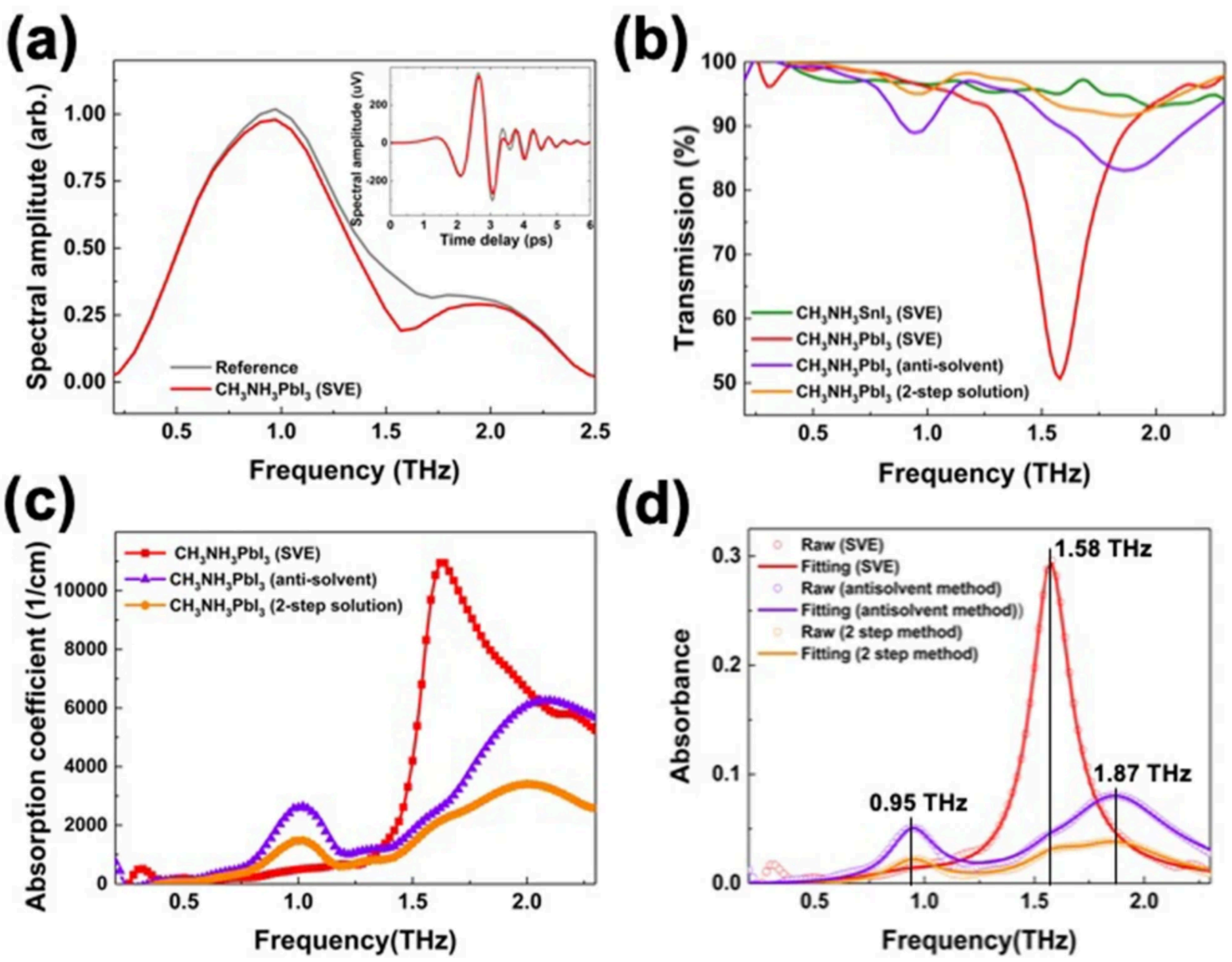

**Figure 1.** THz characterization of $MAPbI_3$ thin films by different fabrication methods. (a) THz waveforms and (b) transmission spectra: the SVE film uniquely shows ~50 % suppression at 1.58 THz. (c) Absorption coefficient spectra (>11,000 $cm^{-1}$ at 1.58 THz for SVE). (d) THz absorbance decomposed into three Lorentz oscillators at 0.95, 1.58, and 1.87 THz. Reproduced from Maeng *et al*., Sci. Rep. 2019, 9, 5811 [21].

DFT phonon calculations on tetragonal $CH_3NH_3PbI_3$ reproduce the 0.95 and 1.87 THz intrinsic modes but not the 1.58 THz peak, directly pointing to a defect-related origin. XPS characterization confirms higher $CH_3NH_2$ molecular density at grain boundaries in SVE films: C and N 1*s* core-level spectra show elevated intensities at 287.0 eV and 402.7 eV ($CH_3NH_2^+$ binding energies), absent in anti-solvent films. SEM reveals smaller grains and higher GB density in SVE films (RMS roughness 18.8 nm vs. 8.4 nm). (Figure 2) The 1.58 THz mode is thus assigned to vibration of the Pb–I bond structurally deformed by the co-incorporated $CH_3NH_2$ molecular defect at GB sites [11,15].

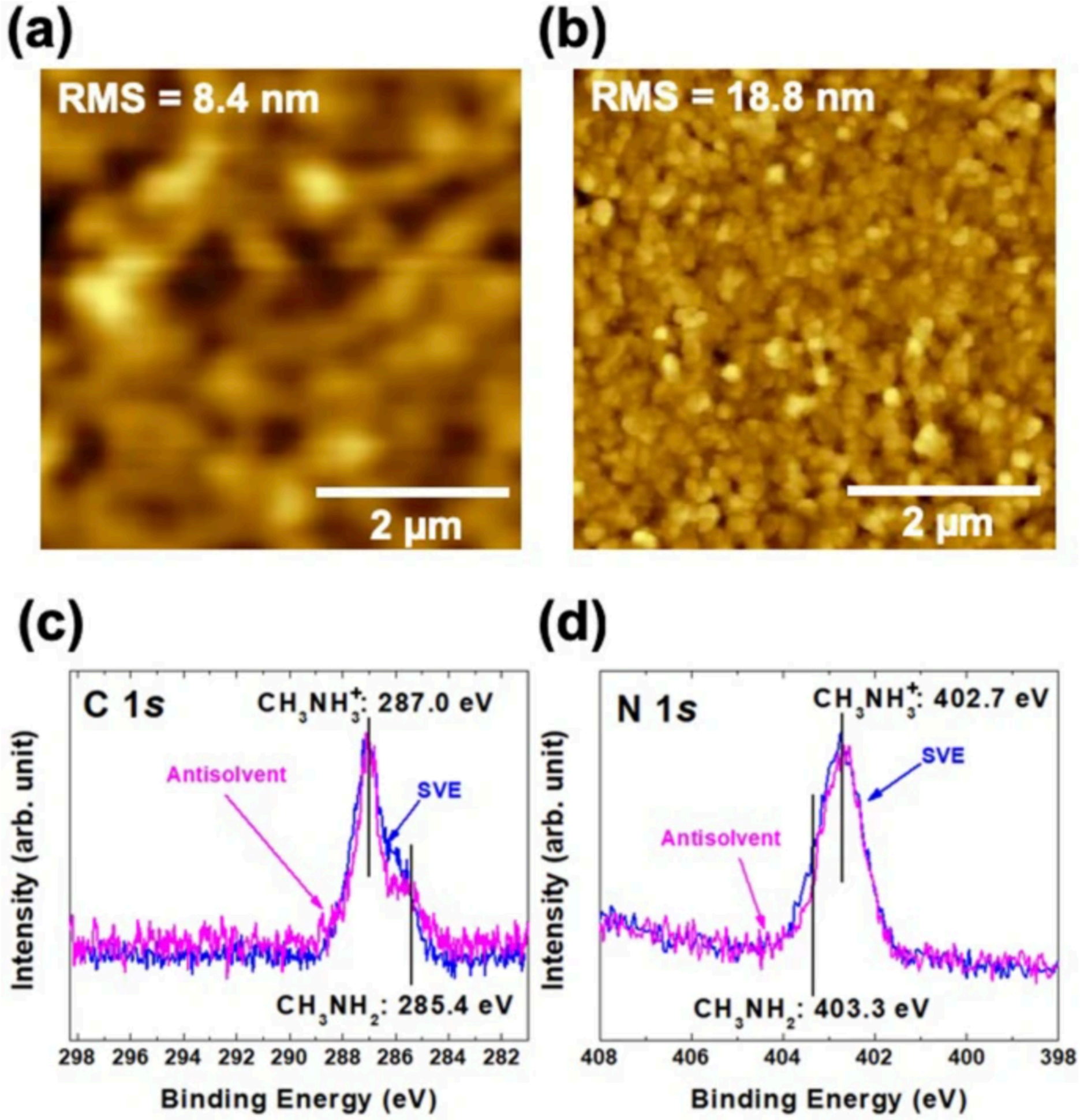

**Figure 2.** XPS and surface morphology evidence for $CH_3NH_2$ grain-boundary defects. (a,b) SEM images: SVE (RMS = 18.8 nm) vs. anti-solvent film (RMS = 8.4 nm). (c,d) C and N 1*s* XPS spectra confirming elevated $CH_3NH_2^+$ intensities at 287.0 eV and 402.7 eV in SVE films. Reproduced from Maeng *et al*., Sci. Rep. 2019, 9, 5811 [21].

### 4.2. Quantitative Correlation: THz Oscillator Strength as a Direct Defect Density Meter

Maeng *et al*. (2020) [27] established the quantitative value of the 1.58 THz fingerprint by systematically varying $CH_3NH_2$ defect density in $MAPbI_3$ through post-annealing under $N_2$ atmosphere and correlating results with XPS-derived defect concentrations. Four $MAPbI_3$ samples (A–D) were prepared at different annealing conditions: no anneal (A), 110 °C/45 min (B, optimal), 150 °C/10 min (C), and 150 °C/30 min (D, over-annealed).

Figure 3 shows the curve-fitting results of C 1*s* core-level spectra and extracted $CH_3NH_2$ relative intensity across samples A–D. The $CH_3NH_2$ defect concentration follows a non-monotonic trend: high in sample A (~18%), nearly eliminated in sample B (~1%), then recovered in C and D (~15%) as over-annealing decomposes the perovskite. LHO fitting yields oscillator strengths at 0.97, 1.62, and 2.01 THz. Only the 1.62 THz oscillator strength reproduces the same non-monotonic pattern—in direct, linear proportion to XPS-quantified $CH_3NH_2$ defect concentration. The 0.97 and 2.01 THz modes remain constant, confirming their intrinsic phonon origin.

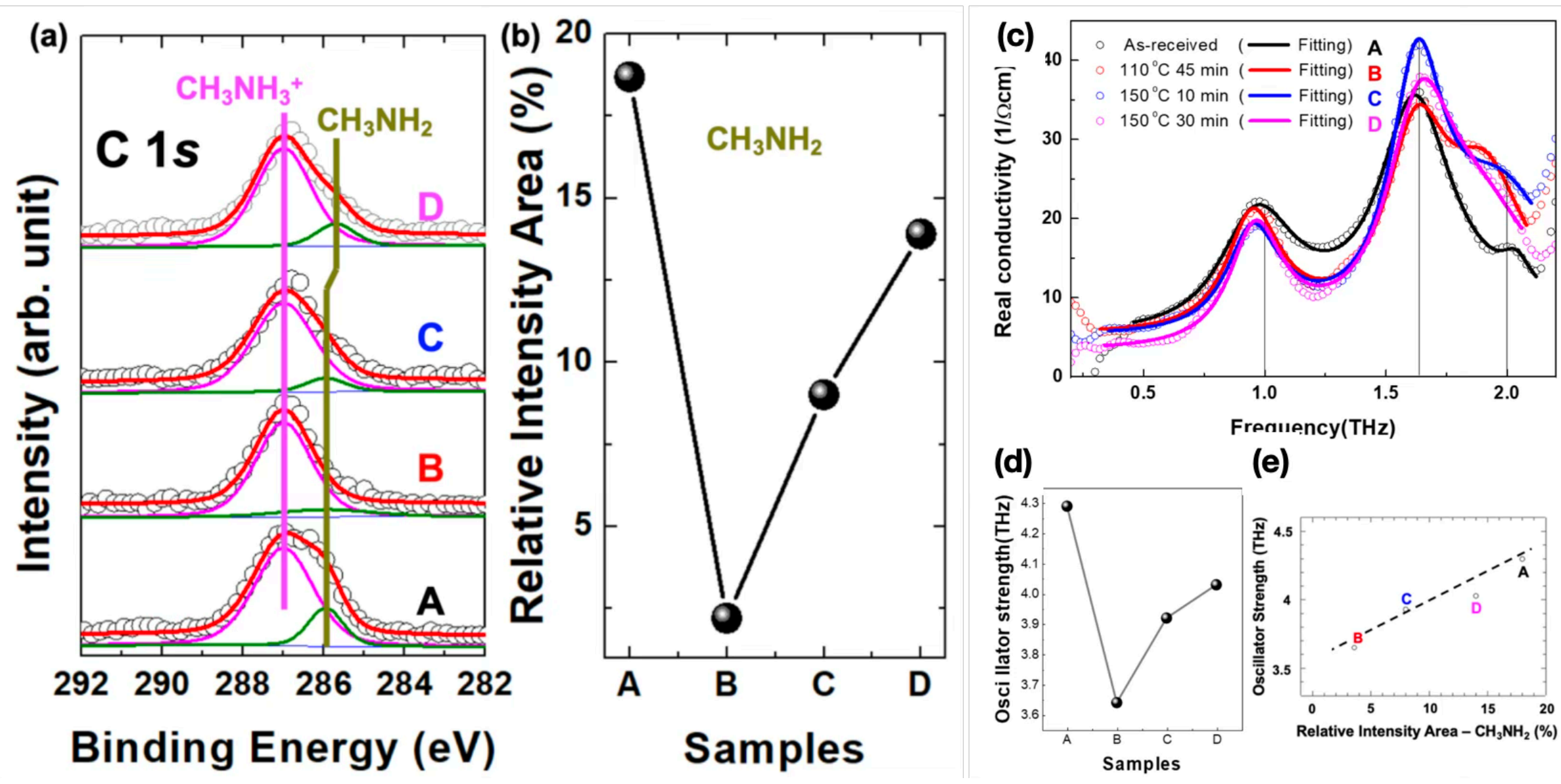

**Figure 3.** Quantitative defect fingerprinting in $MAPbI_3$. (a) the curve fittings of C 1*s* core-level spectra for samples A–D. (b) Relative $CH_3NH_2$ peak intensity (%) vs. annealing condition: non-monotonic behavior linearly correlated with THz oscillator strength at 1.6 THz. Reproduced from Maeng *et al*., Nanomaterials 2020, 10, 721 [27].

This linear THz oscillator strength–defect concentration relationship establishes THz-TDS as a quantitative, contact-free defect meter that bypasses the surface-depth limitation of XPS and the species-averaging limitation of PL [12]. Eliminating this defect (Sample B, $\Omega$(1.6 THz) ≈ 0) directly correlates with recovery of full OHP crystal structure and optimal device performance. This constitutes Pillar I of the three-pillar framework.

### 4.3. $MAPbBr_3$: Defect-Immune Intrinsic Phonon Fingerprints

Maeng *et al*. (2020) [25] demonstrated that SVE-fabricated $MAPbBr_3$ shows three sharp, stable phonon absorptions at 0.8, 1.4, and 2.0 THz completely insensitive to grain-boundary $CH_3NH_2$ defect chemistry. Despite XPS confirming substantial $CH_3NH_2$ reduction upon post-annealing, the THz phonon modes remain unchanged in frequency, oscillator strength, and linewidth (Figure 4). This contrast with $MAPbI_3$ is mechanistically explained by the stronger Pb–Br bond, which resists structural deformation upon $CH_3NH_2$ incorporation and thus does not produce a defect-induced vibration shift.

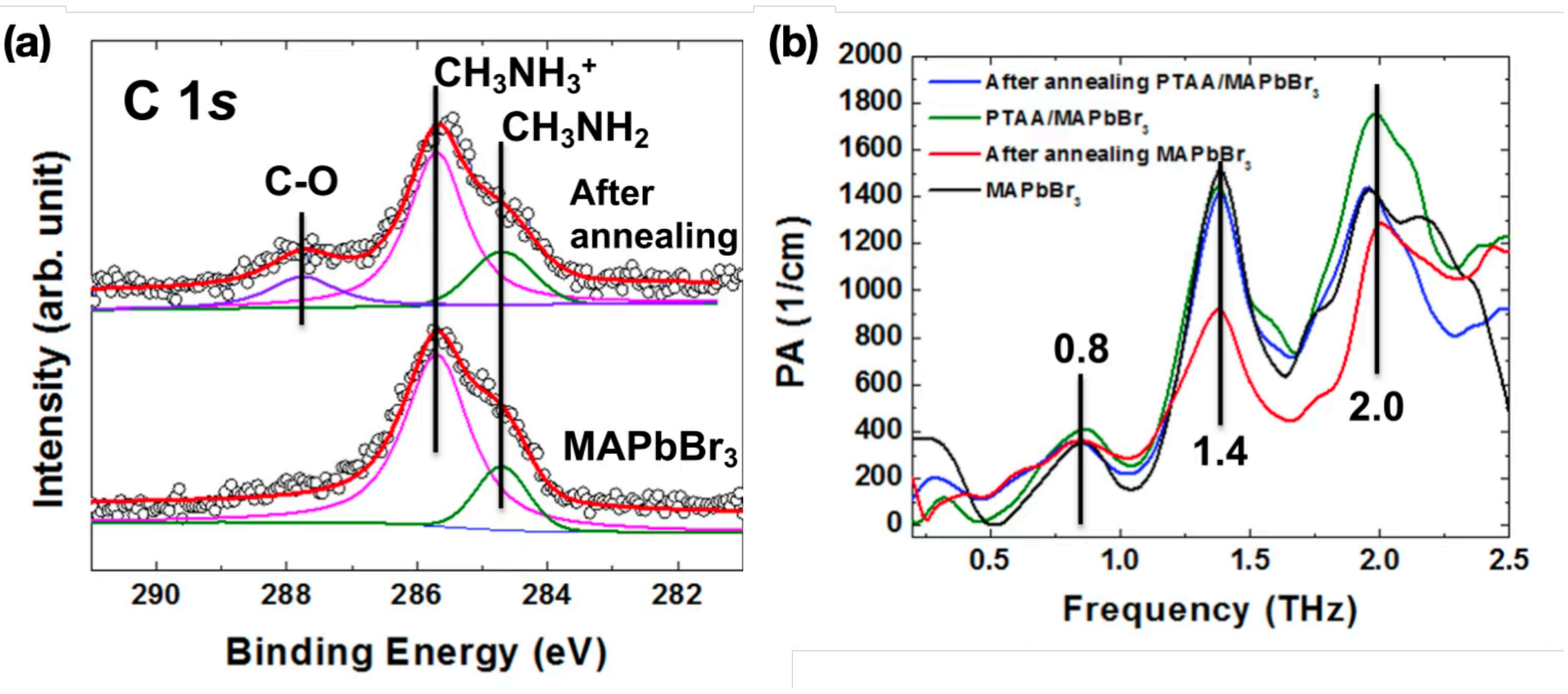

**Figure 4.** Defect-immune THz phonon fingerprints of SVE-fabricated $MAPbBr_3$. (a) C 1*s* core-level spectra before/after 150 °C annealing: $CH_3NH_2$ molecular defect (287.0 eV) substantially eliminated. (b) THz absorption spectra: three phonon modes at 0.8, 1.4, and 2.0 THz unchanged by defect chemistry modification. Reproduced from Maeng *et al*., NPG Asia Mater. 2020, 12, 53 [25].

First-principles phonon calculations using the temperature-dependent effective potential (TDEP) method on cubic $MAPbBr_3$ at 300 K (Figure 5) [25] assign the three modes as: 0.8 THz—transverse libration of the MA cation; 1.4 THz—longitudinal optical vibration of the Pb–Br framework; 2.0 THz—self-vibration of the Br sublattice. This DFT-level assignment provides a rigorous quantum-mechanical basis for the THz fingerprint assignments.

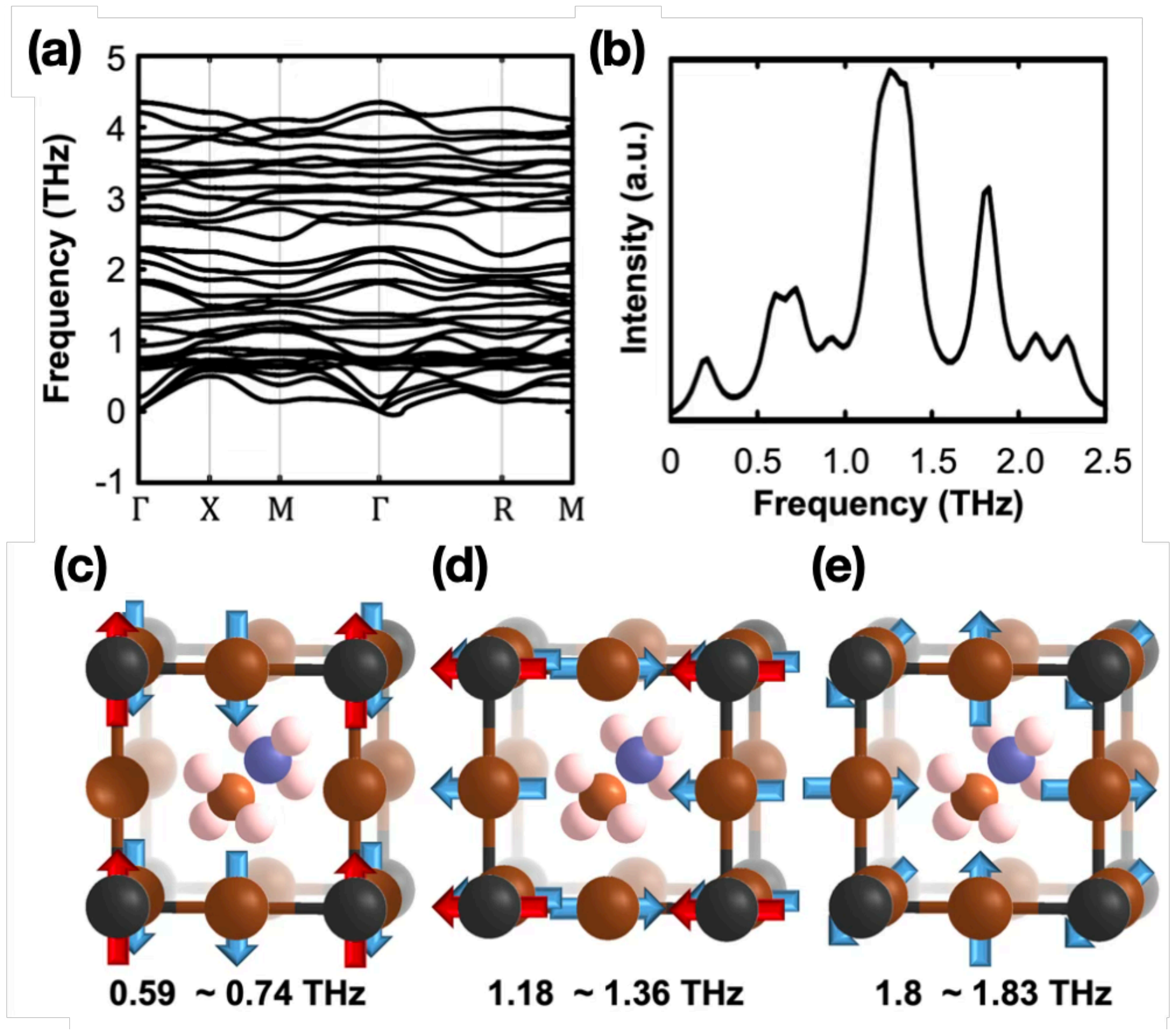

**Figure 5.** TDEP-DFT phonon analysis of $MAPbBr_3$. (a) Phonon dispersion at 300 K with three IR-active branches. (b) Simulated IR absorption vs. experimental THz conductivity. (c–e) Phonon eigenvectors at 0.8, 1.4, and 2.0 THz. Reproduced from Maeng *et al*., NPG Asia Mater. 2020, 12, 53 [25].

$MAPbBr_3$ thus represents the 'clean phonon' regime in the OHP THz fingerprint library: when the THz spectrum matches the DFT-predicted intrinsic phonon structure without additional oscillators, the grain boundaries harbor no THz-active molecular defect modes—a spectroscopic certificate of GB vibrational integrity. This does not exclude the possible presence of THz-silent defect states detectable only by complementary techniques.

### 4.4. $FAPbI_3$: Phase-Boundary Phonon Fingerprints

Phase control is a central challenge for FA-based perovskites: the photoactive α-$FAPbI_3$ (black cubic phase) is thermodynamically metastable at room temperature and reverts to the photoinactive δ-phase (yellow hexagonal phase) unless stabilized. THz-TDS provides a direct, non-destructive window into phase-boundary formation that complements XRD, offering sensitivity to the interfacial bonding state rather than the bulk crystal structure.

Maeng *et al*. (2021) [22] demonstrated that δ/α-mixed-phase $FAPbI_3$ single-crystal surface regions, formed by controlled annealing at 140°C for 30 min, exhibit two unique THz absorption peaks at 2.0 and 2.2 THz absent from both pure δ- and α-phases (Figure 6). The pure α-phase (180°C) shows a single phonon mode at ~1.45 THz. Phase assignments were confirmed by Pb 4*f* and I 4*d* core-level spectra, which show no new chemical states at the mixed-phase boundary.

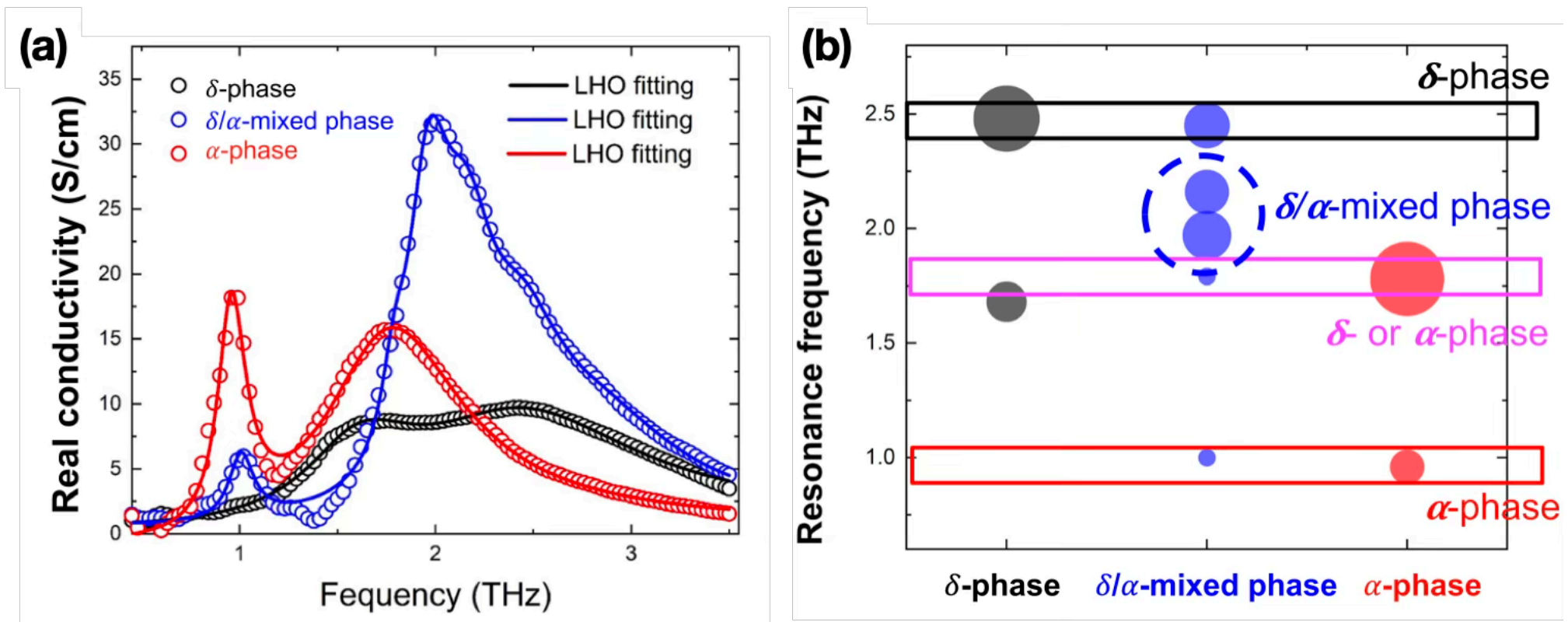

**Figure 6.** Phase-sensitive THz fingerprints in $FAPbI_3$. (a–c) Surface morphology: δ-phase (as-deposited), δ/α-mixed (140°C/30 min), α-phase (180°C/30 min). (d) THz real conductivity: mixed phase uniquely shows 2.0 and 2.2 THz modes. (e,f) Pb 4f and I 4d XPS confirming phase identity. Reproduced from Maeng *et al*., NPG Asia Mater. 2021, 13, 75 [22].

Molecular dynamics (MD) simulations—employed here instead of static DFT to capture the thermal disorder inherent to the δ/α mixed-phase interface at finite temperature—confirm that the 2.0 and 2.2 THz modes originate from compressed Pb–I bond lengths at the structurally frustrated δ/α interface layer. These interfacial phonon modes are: (i) upshifted from the bulk α-phase 1.45 THz mode by bond compression, (ii) invisible to XRD, and (iii) uniquely correlated with the presence of a seamless phase boundary. They constitute a THz fingerprint of phase-boundary structural integrity.

In mixed-halide $FAPb(Br_xI_{1-x})_3$, the THz fingerprint is tunable by composition [23]. I-rich formulations ($x \leq 0.3$) show strong absorption at ~1.6 THz from I-rich GB domains, while Br-rich compositions ($x \geq 0.5$) transition toward three-phonon behavior (Figure 7). This composition–THz fingerprint map directly guides halide optimization for minimizing I-related GB defect density in FA-based tandem cells [24].

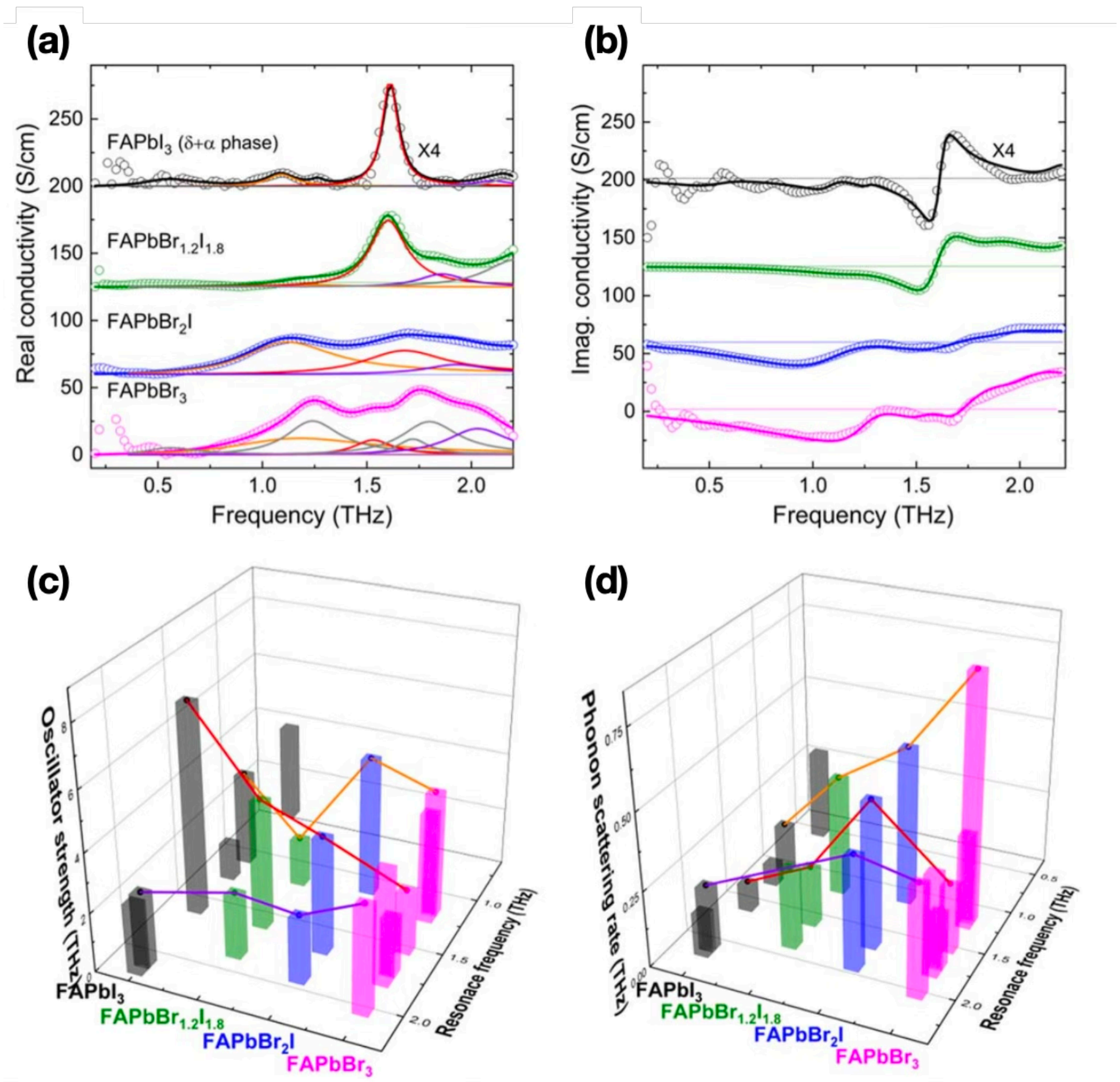

**Figure 7.** Composition-dependent THz fingerprints in $FAPb(Br_xI_{1-x})_3$ thin films by SVE. THz conductivity spectra across the composition series: I-rich compositions show large ~1.6 THz GB absorption; Br-rich compositions converge on three intrinsic phonon modes. Reproduced from Maeng *et al*., Appl. Phys. Express 2021, 14, 121002 [23].

#### 4.5. γ-$CsPbI_3$: Grain-Size-Independent Phonon Fingerprints Confirming Defect-Free GBs

The all-inorganic γ-$CsPbI_3$ (orthorhombic, black phase) is thermally stable compared to organic-cation analogs. The absence of $CH_3NH_2$ decomposition products at grain boundaries provides a direct test of whether the 1.58 THz defect fingerprint in SVE-fabricated films is uniquely tied to organic MA cation chemistry.

Maeng *et al*. (2023) [28] fabricated γ-$CsPbI_3$ films by SVE with three grain sizes (A: 300 nm, B: 350 nm, C: 460 nm) and measured THz conductivity (Figure 8). All three samples show three clean phonon absorption peaks at 0.9, 1.5, and 1.8 THz with no additional absorption. Both resonance frequencies and oscillator strengths are entirely grain-size-independent. DFT phonon calculations assign the three modes as: 0.9 THz (Cs rattling mode), 1.5 THz (Pb–I–Pb bending), and 1.8 THz (Pb–I stretching).

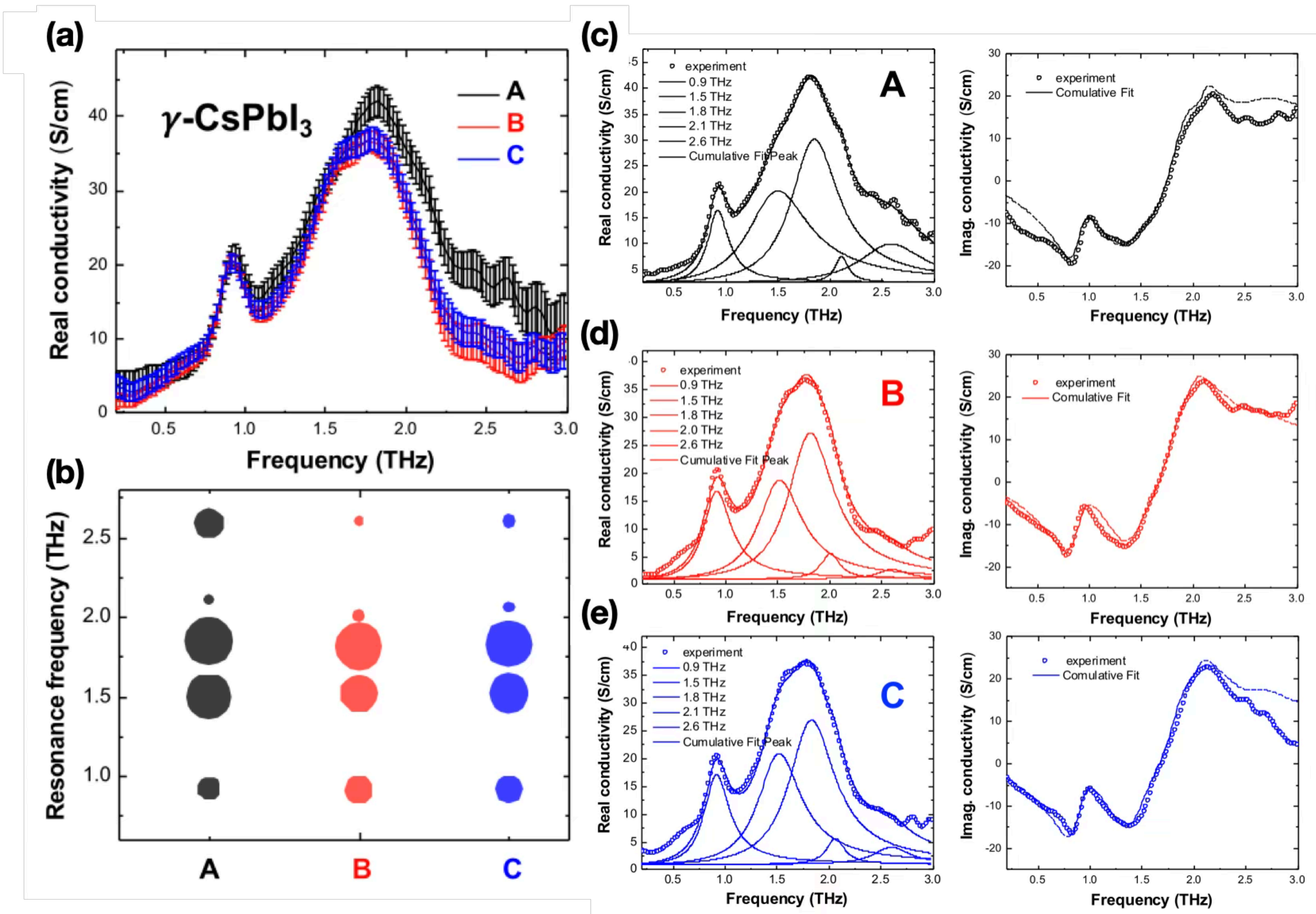

**Figure 8.** Grain-size-independent THz phonon fingerprints in $\gamma$-$CsPbI_3$ thin films. (a) Real conductivity for samples A (300 nm), B (350 nm), C (460 nm): identical three-phonon spectra. (b) LHO resonance frequencies and oscillator strengths vs. grain size: no dependence observed. (c–e) Individual LHO component fits for A, B, C. Reproduced from Maeng *et al*., Mater. Today Phys. 2023, 30, 100960 [28].

The complete absence of grain-size-dependent THz absorption in $\gamma$-$CsPbI_3$ confirms: (i) SVE does not introduce molecular defects at $CsPbI_3$ grain boundaries; and (ii) the THz fingerprint is a reliable, grain-size-independent indicator of GB chemical integrity. The stark contrast between the defect-active 1.58 THz mode in $MAPbI_3$ and the defect-free three-phonon spectrum of $\gamma$-$CsPbI_3$ provides compelling evidence that the 1.58 THz fingerprint is a molecular-chemistry-specific signal, not a structural grain-boundary artifact.

## 5. The Three-Pillar THz Defect Engineering Framework

The THz fingerprint library presented above directly enables a three-pillar engineering framework that links spectroscopic defect identification to process optimization and device performance enhancement. Figure 9 illustrates the interconnected cycle: Pillar I provides quantitative measurement capability, Pillar II supplies compositional context, and Pillar III closes the engineering loop through targeted defect elimination.

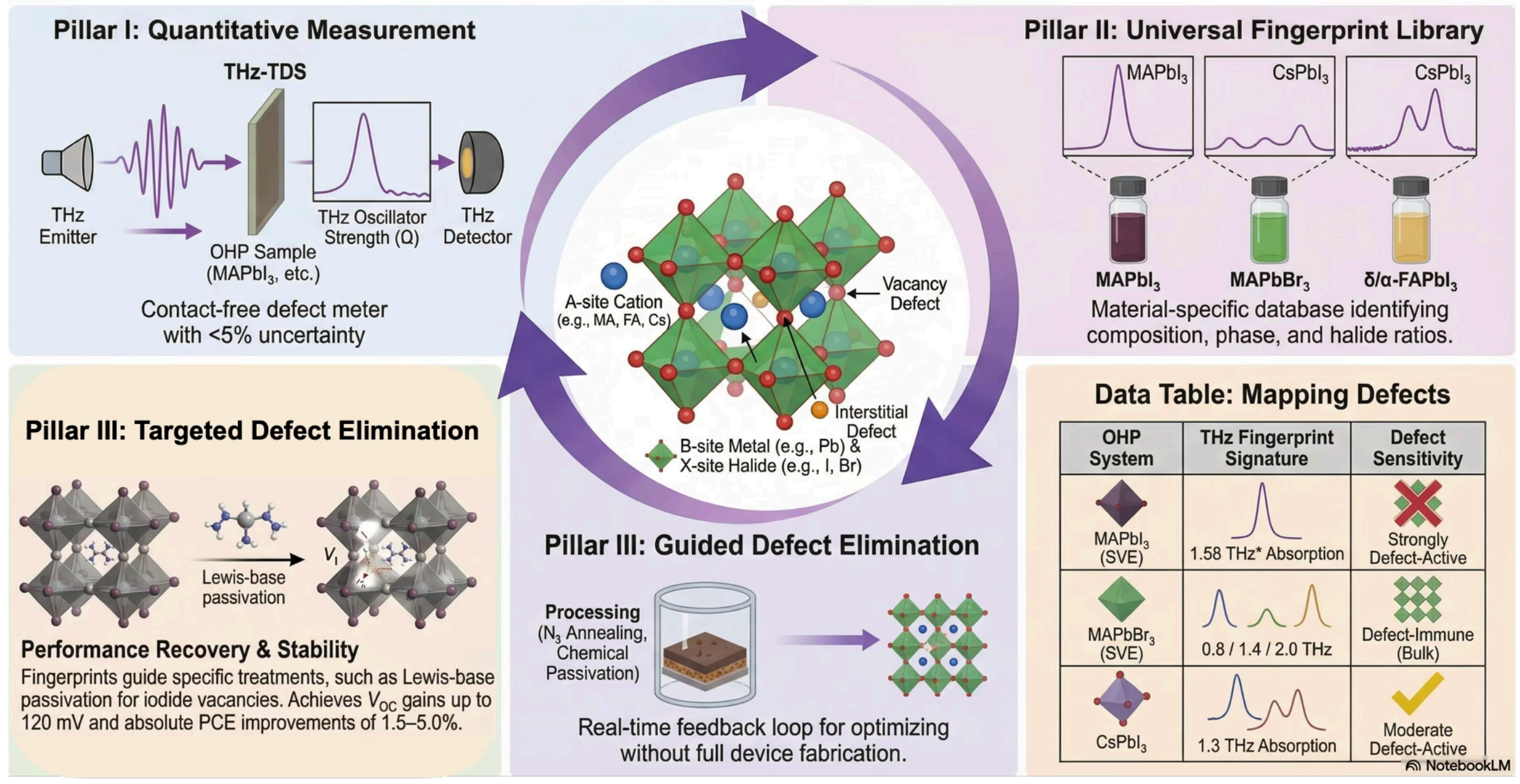

**Figure 9.** Schematic of the three-pillar THz defect engineering framework. Three interconnected pillars—(I) quantitative defect measurement, (II) universal fingerprint library, and (III) fingerprint-guided defect elimination—operate in a continuous cycle around the central THz-TDS engine, converting raw spectral data into actionable process feedback for OHP solar cell manufacturing.

**Table 1.** THz defect fingerprint library for SVE-fabricated OHP thin films. Asterisk (*) denotes defect- or phase-boundary-specific modes.

| OHP System | THz Modes (THz) | Defect Sensitivity | Physical Assignment | Ref. |
|---|---|---|---|---|
| $MAPbI_3$ (SVE) | 0.95/1.58*/1.87 | Strongly defect-active (1.58 THz) | *$CH_3NH_2$ mol. defect at GB; others = bulk phonons | [21],[27] |
| $MAPbBr_3$ (SVE) | 0.8/1.4/2.0 | Defect-immune | All intrinsic phonons (TDEP-DFT verified) | [25] |
| δ/α-$FAPbI_3$ | 1.45/2.0*/2.2* | Phase-boundary sensitive | *Interfacial phonons at seamless δ/α boundary | [22] |
| $FAPb(Br_xI_{1-x})_3$ | Composition-dep. | Halide-composition-tunable | I-rich: GB absorption; Br-rich: intrinsic phonons | [23] |
| γ-$CsPbI_3$ (SVE) | 0.9/1.5/1.8 | Grain-size independent | Bulk phonons; defect-free GB confirmed | [28] |

### 5.1. Pillar I—Quantitative Defect Measurement

The linear correlation between Ω(1.6 THz) and XPS-quantified $CH_3NH_2$ concentration [27] transforms THz oscillator strength into a non-contact defect concentration meter. The measurement uncertainty, estimated from repeated LHO fits on identically prepared films, is <5% for films >100 nm [27]; spectral isolation of the 1.6 THz mode from the 0.97 and 2.0 THz phonons enables unambiguous LHO decomposition; and room-temperature ambient measurement eliminates the vacuum-sample-transfer requirement of XPS. This enables: (i) substrate-by-substrate, process-step-resolved defect quantification; (ii) inline monitoring of post-annealing conditions without device fabrication; and (iii) tracking of defect density evolution across roll-to-roll SVE production sequences.

### 5.2. Pillar II—Universal THz Fingerprint Library

The four OHP systems reviewed define a material-specific THz fingerprint library (Table 1) spanning defect-active, defect-immune, phase-sensitive, and composition-tunable regimes. This library enables unambiguous spectroscopic discrimination of: (i) fabrication method (SVE vs. solution-processed, through presence/absence of 1.58 THz mode); (ii) OHP composition (distinct phonon frequencies for MA, FA, Cs); (iii) crystallographic phase (δ vs. α vs. δ/α-mixed through 1.45 vs. 2.0/2.2 THz modes); and (iv) halide composition (THz oscillator strength map for $FAPb(Br,I)_3$ series). No single conventional technique provides this combination of information non-destructively from a single measurement.

### 5.3. Pillar III—Fingerprint-Guided Defect Elimination

The THz fingerprint provides immediate, quantitative feedback for elimination strategies. Thermal post-annealing at 110°C for 45 min in $N_2$ reduces Ω(1.6 THz) from ~0.37 to near zero in $MAPbI_3$ [27]. Electric-bias treatment (15 V) of $MAPbBr_3$ single-crystal surfaces eliminates 55% of $CH_3NH_2$ through field-driven ionic migration [29]. Ultra-thin P3HT (7 nm) HTL deposition creates a chemically clean interface and provides 2-week air stability protection [30]. Advanced sequential ligand strategies such as the two-step dedicated-ligand approach by Choy *et al*. [31]—applying DFABr for grain boundaries and $4APPBr_2$ for grain surfaces—have demonstrated >93% PCE retention after 3,000 hours under high humidity; THz defect fingerprinting before and after such treatments directly quantifies elimination efficacy, providing a spectroscopic certificate of passivation.

**Table 2.** The three-pillar THz defect engineering framework: capabilities, evidence base, and application scope.

| Pillar | Core Capability | Key Evidence | Application |
|---|---|---|---|
| I | THz oscillator strength Ω(1.6 THz) as contact-free defect meter (<5% uncertainty) | Linear Ω–[$CH_3NH_2$] correlation in $MAPbI_3$ (samples A–D) [27] | Substrate-by-substrate process monitoring; inline SVE QC without device fabrication |
| II | Material-specific THz fingerprint database spanning 4 OHP systems | Defect-active ($MAPbI_3$), immune ($MAPbBr_3$), phase-sensitive ($FAPbI_3$), tunable ($FAPb(Br,I)_3$), inorganic-clean (γ-$CsPbI_3$) [21]–[28] | Non-contact discrimination of composition, phase, and halide ratio in a single scan |
| III | THz fingerprint as real-time feedback loop for defect elimination | $N_2$ anneal (110°C/45 min): Ω→0 [27]; E-bias: 55% $CH_3NH_2$ eliminated [29]; P3HT HTL: 2-week stability [30] | Iterative passivation optimization without full device fabrication |

## 6. Discussion

The THz defect fingerprint approach occupies a unique position in the OHP characterization landscape. Early OPTP-THz studies focused on carrier mobility and dynamics [14], establishing THz-TDS as a benchmark for transport quality but not defect identity. Temperature-dependent Time-Resolved Terahertz Spectroscopy (TRTS) by Milot *et al*. [32] demonstrated how charge-carrier recombination and mobility in $MAPbI_3$ evolve across phase transitions. Transmission THz-TDS studies by La-o-vorakiat *et al*. [11] and Sendner *et al*. [15] mapped intrinsic phonon modes and their coupling to carrier transport. Near-field THz nanoimaging by Kim *et al*. [17] spatially resolved GB trap filling with < 20 nm resolution—a nanoscale complement to the far-field oscillator strength approach. Hempel *et al*. [33] demonstrated that THz and microwave transient measurements alone can quantitatively predict PCE of perovskite solar cells, bridging the gap between non-contact spectroscopy and device-level benchmarking.

Table 3 compares THz-TDS against principal competing characterization techniques. Compared to PL: THz oscillator strength is immune to reabsorption, surface passivation artifacts, and excitation-dependent injection. Compared to impedance spectroscopy: THz-TDS requires no contacts and accesses THz-frequency lattice dynamics inaccessible to kHz–MHz impedance measurements. Compared to XPS: THz-TDS is non-destructive, non-contact, operates under ambient conditions, and probes the entire film depth.

**Table 3.** Comparative capabilities of THz-TDS vs. conventional OHP defect characterization methods. ✓ = capable, ✗ = not capable, Partial = limited.

| **Criterion** | **THz-TDS** | **XPS** | **PL** | **Impedance** | **XRD/SEM** |
|---|---|---|---|---|---|
| Defect species ID | ✓ Direct | ✓ Direct | ✗ Indirect | ✗ Indirect | ✗ None |
| Non-contact | ✓ | ✗ UHV | ✓ | ✗ Contacts | ✓ |
| Non-destructive | ✓ | Partial | ✓ | Partial | ✓ |
| Quantitative | ✓ Linear | Partial | Partial | Partial | ✗ |
| Ambient/RT | ✓ | ✗ | ✓ | Partial | ✓ |
| Full-depth probe | ✓ | ✗ ~10 nm | ✓ | ✓ | ✓ |
| Phase sensitivity | ✓ Interfacial | ✗ | Limited | ✗ | ✓ Bulk |

The distinct response of the 1.58 THz mode in $MAPbI_3$ compared to the defect-immune nature of $MAPbBr_3$ provides a critical physical insight into the lattice-defect interaction. [21, 25] While both systems were fabricated via SVE, the contrast originates from the disparity in Pb–X (X = I, Br) bond energies. The stronger ionic bonding and higher lattice rigidity of the bromide framework suppress structural deformations induced by incorporated $CH_3NH_2$ molecular defects. [25] In contrast, the relatively soft Pb–I lattice in $MAPbI_3$ is more susceptible to local distortions, which activate the specific vibrational mode observed at 1.58 THz. [21] This highlights that THz-TDS does not merely detect the presence of defects but probes the mechanical coupling between the defect species and the host lattice, offering a unique spectroscopic gauge for lattice stability.

Phase segregation under illumination in mixed-halide OHPs—documented by Liu *et al*. [24] and others—is directly observable in the THz composition-fingerprint map: I-rich GB domains correspond to elevated Ω(1.6 THz), while Br-rich compositions maintain phonon-dominated spectra. Sun *et al*. [34] recently demonstrated real-time detection of $MAPbI_3$ aging via monotonic decay of a THz absorption peak at 0.968 THz, tracking degradation onset non-destructively—consistent with the fingerprint-guided monitoring concept presented here.

A potential concern regarding the linear correlation between surface-sensitive XPS and bulk-probing THz oscillator strength is addressed by the inherent nature of the SVE process. Unlike solution-processing where solvent evaporation can lead to vertical defect gradients, sequential vacuum evaporation facilitates a more homogeneous distribution of trapped molecular entities throughout the film thickness. [27] Consequently, the surface $CH_3NH_2$ concentration serves as a reliable proxy for the total defect density within the grain boundary network, validating the use of the 1.6 THz oscillator strength as a non-destructive bulk defect meter for manufacturing-scale quality control. [27]

Looking ahead, critical open questions include: (i) the quantitative THz oscillator strength–PCE calibration curve that would transform THz-TDS into a direct device performance predictor; (ii) extension of the fingerprint library to ternary/quaternary compositions and 2D Ruddlesden–Popper phases; (iii) time-resolved THz-TDS under pulsed illumination to distinguish photo-generated from thermally stable defects; and (iv) integration of THz measurement heads into inline SVE deposition systems for real-time manufacturing quality control.

## 7. Conclusions

OHP solar cells now exceed 26.7% PCE in certified single-junction devices and 34.85 % in perovskite–silicon tandems, yet grain-boundary defects remain the principal barrier to bridging the gap with the Shockley–Queisser limit. This short review has demonstrated that THz-TDS fulfills the role of a composition-resolved defect fingerprint tool through a library spanning four archetypal OHP systems:

(1) SVE-fabricated $MAPbI_3$ carries a unique 1.58 THz $CH_3NH_2$ molecular defect fingerprint ($\alpha > 11,000$ cm$^{-1}$) [21]. The oscillator strength scales linearly with XPS-quantified defect concentration [27], establishing THz-TDS as a quantitative, contact-free defect meter.

(2) $MAPbBr_3$ shows DFT-verified defect-immune THz phonon modes at 0.8, 1.4, and 2.0 THz, defining the 'clean grain boundary' spectroscopic benchmark for bromide-based compositions [25].

(3) δ/α-mixed $FAPbI_3$ displays interfacial phonon modes at 2.0 and 2.2 THz from seamless phase boundaries—invisible to XRD but accessible by THz—enabling non-contact phase-boundary quality assessment [22].

(4) $FAPb(Br,I)_3$ provides a composition-tunable THz map from defect-active (I-rich, ~1.6 THz) to defect-immune (Br-rich, phonon-dominated), directly guiding halide ratio optimization [23].

(5) γ-$CsPbI_3$ shows grain-size-independent bulk phonon modes at 0.9, 1.5, and 1.8 THz, confirming defect-free grain boundaries in all-inorganic perovskite absent of organic cation chemistry [28].

Looking forward, the establishment of a direct 'THz Fingerprint–to–PCE' calibration curve will be the final step in transitioning THz-TDS from a laboratory diagnostic tool to a predictive manufacturing engine. [33] By integrating THz measurement heads directly into inline SVE deposition systems, real-time feedback loops can be realized to optimize passivation strategies without the need for full device fabrication. This shift toward spectroscopy-guided manufacturing offers a scalable and cost-effective pathway to close the remaining efficiency gap with the Shockley–Queisser limit, ensuring both the high performance and long-term reliability required for the commercialization of perovskite photovoltaics.

**Author Contributions:** Conceptualization, I.M. and M.-C.J.; Writing—Original Draft Preparation, I.M. and Y.M.L.; Writing—Review & Editing, S.J.O. and M.-C.J.; Supervision, I.M. and M.-C.J. All authors have read and agreed to the published version of the manuscript.

**Funding:** This work was supported by

**Institutional Review Board Statement:** Not applicable.

**Informed Consent Statement:** Not applicable.

**Data Availability Statement:** The raw data supporting the conclusions of this article will be made available by the authors on request.

**Conflicts of Interest:** The authors declare no conflicts of interest. The funders had no role in the design of the study; in the collection, analyses, or interpretation of data; in the writing of the manuscript; or in the decision to publish the results.